\documentstyle[aps,prl,twocolumn,epsf]{revtex}
 
\begin{document}

{\bf Comment on "Density of states and reflectionless tunneling in NS junction
with a barrier"}

\begin{abstract}
In a recent paper Schechter, Imry and Levinson (cond-mat/9709248) \cite{r1} analysed
the density-of-states (DOS) and the conductance of a superconductor-normal metal
(S/N) junction with a barrier. They showed that the DOS in the N conductor near the barrier
is strongly affected by the superconductor if the energy $\epsilon$ is small:
$l_{n} \sqrt{\epsilon/D}{\ll} \Gamma {\ll} 1$, where $l_{n}$ is the mean free path, $D$ is the 
diffusion constant, and $\Gamma$ is the barrier transmittance. We would like to note that
this problem was analysed in several papers with the use of different approaches. In particular,
the author has calculated the DOS and the conductance of S/N structures with a barrier
using quasiclassical matrix Green's function technique well developed in the theory 
of superconductivity \cite{r2}. It was shown that in an one-dimensional S/N structure a pseudogap appears
in the N conductor near the S/N interface, and the DOS approaches it's normal value  
at energies $\epsilon$ exceeding a characteristic energy $\epsilon_{N}$ 
related to the interface transmittance; here $\epsilon_{N}=D/(2 R_{b} \sigma)^{2}$, $R_{b}$ 
is the interface resistance per unit area, and $\sigma$ is the specific conductivity of the N 
conductor. In a planar S/N junction a real gap $\epsilon_{N}$ appears in the excitation spectrum
of the N electrode. It was shown that peculiarities in the DOS at small energies lead to a subgap
conductance at zero temperature. This subgap conductance is caused by a component of the
current which is known in the theory of the Josephson effect as the interference component of
the Josephson current. An applied magnetic field suppresses the subgap conductance \cite{r3}.
Therefore, the results for the DOS and the conductance in Ref \cite{r1} are similar to the ones
obtained earlier, but their interpretation of the subgap conductance differs from that given in
Ref.\cite{r2} because they use another approach (Landauer formula) for the calculation of the conductance 
.

\end{abstract}

In the last years many papers have been published in which the conductance of mesoscopic
S/N structures was studied. Two approaches were used in these works. One of them is based 
on the calculation of a scattering matrix and on the Landauer formula. The other employs
the method of matrix quasiclassical Green's functions well developed in the theory of 
superconductivity (see review articles \cite{r4,r5}). In Ref.\cite{r2} the DOS and 
the conductance of S/N junctions have been calculated on the basis of the Green's function
technique. Both one-dimensional and planar structures have been analysed under assumption that 
a barrier exists at the interface. It was shown that in the case of a one-dimensional 
S/N structure a pseudogap appears in the N conductor near the S/N interface, i.e. the DOS 
increases  with increasing energy $\epsilon$ and approaches it's value in the normal state
(in the absence of the superconductor) at energies much larger than $\epsilon_{N}$.
It was supposed also that the barrier strength is high enough so that 
the condition $\epsilon_{N}{\ll}\Delta$ is satisfied. This result agrees with that obtained 
in Ref.\cite{r1}. In planar S/N junction the DOS is zero in the N electrode at energies
$\epsilon < \epsilon_{N}$ (a real energy gap). Using equations for matrix quasiclassical
Green's functions with boundary conditions at the S/N interface \onlinecite{r6,r7,r8}, we have 
calculated the conductance of these structures and showed that a subgap conductance appears
at zero temperature. The differential conductance $dI/dV$ has a peak at zero bias in a one-dimensional 
S/N structure and the width of this peak is about $\epsilon_{N}/e$. In the case of a planar S/N 
structure this peak is located at $\epsilon_{N}/e$. The subgap conductance and a dip in the DOS
at small energies are related to an anomalous proximity effect at small energies (perhaps  
Zaitsev was the first who obtained the subgap conductance in a short S/N/N' structure  
\onlinecite{r9,r10}). The condensate amplitude $F_{N}$ in the N conductor near the interface is of 
the order of the condensate amplitude in the superconductor if the energy is small enough:  
$\epsilon < \epsilon_{N}$. In the Green's function technique the appearance of the subgap  
conductance may be explained as a contribution of a component of the current which is 
proportional to $F_{N}$ and is known in the theory of the Josephson effect as the 
interference component of the Josephson current. A small magnetic field suppresses the subgap conductance
\cite{r3,r10}. Therefore, both phenomena (peculiarities in the DOS at small energies in the N conductor 
and the subgap conductance) are connected with each other and are caused by the anomalous proximity effect.
The authors of Ref.\cite{r1} used essentially the same method of the quasiclassical Green's functions for 
the calculation of the DOS (the Usadel equation), but they employed another method (the Landauer formula) 
for the calculation of the conductance. The results of the work \cite{r1} coincide with that obtained in
Ref. \cite{r2}, but the interpretation of the subgap conductance is close to that given in Ref. \cite{r11} 
and is complimentary to ours.

\vspace*{0.1truecm}

Anatoly F.Volkov

Departement of Physics,Lancaster University,

Lancaster LA1 4YB,U.K.

and

the Institute of Radioengineering and

Electronics of the RAS,

Mokhovaya 11, Moscow 103907,Russia

\end{document}